\documentclass[preprint,prd,groupedaddress,nofootinbib,showpacs,rotate]{revtex4}

\begin{document}

\title{Bound states for one-electron atoms in higher dimensions }

\author{Nelson R. F. Braga}
\email{braga@if.ufrj.br}
\affiliation{Instituto de F\'{\i}sica,
Universidade Federal do Rio de Janeiro, Caixa Postal 68528, RJ 21941-972 -- Brazil}
\author{Rafael D'Andrea }
\email{dandrea@if.ufrj.br}
\affiliation{Instituto de F\'{\i}sica,
Universidade Federal do Rio de Janeiro, Caixa Postal 68528, RJ 21941-972 -- Brazil}


\begin{abstract} 
We study the Schr\"odinger equation for one-electron atoms in space-times
with $d \ge 4$ spatial dimensions where the Gauss law is assumed to be valid. 
It is shown that there are no normalizable wave functions corresponding to bound states.
The consistency with the classical limit is discussed.
\end{abstract}


\maketitle

\section{Introduction}

Currently, string theory seems to be the best candidate for a quantum theory of gravity
or possibly for a unified theory of fundamental interactions.
Consistency of string theory, in the standard formulations, requires the existence of
extra spatial dimensions \cite{Polchinski:1998rq,Zwiebach:2004tj}. 
Presently, there is no experimental evidence of such extra dimensions. This indicates 
that, if they exist, they are compactified into small length scales 
still unaccessible to our measurements.

One way to find upper bounds for the ``sizes" of these extra dimensions 
is to consider microscopic physical systems for which we have a very well 
established agreement between theoretical modeling and experimental data.
Then, one looks at the corresponding version of the physical system 
assuming the existence of non-compact extra dimensions at that length scale.
One interesting example is the hydrogen atom (or, more generally, a one-electron atom).

Physical systems subject to Gaussian central forces in 
different spatial dimensions were discussed  a long time ago by
Ehrenfest \cite{Ehrenfest}. 
He showed that a classical planetary system does not have stable orbits for spaces with 
more than three non-compact spatial directions. 
He also discussed the Bohr atomic model in such spaces. 

The quantum mechanical spectrum of a one electron atom in higher dimensional spaces, 
assuming the Gauss law to hold, was discussed in ref. \cite{Mostepanenko}.
There, it was shown that there is no lower bound for negative energy solutions.
As a consequence, there would be no stable atom in such spaces.

Many recent articles (see, e.g.,  \cite{Burgbacher,SUSYatom}) 
disregard Gauss law when discussing the hydrogen atom or its supersymmetric 
extension in higher dimensional spaces, taking the eletrostatic potential to 
fall as $1/r$. 
In string theory, such approach is consistent with the idea of some brane-world 
scenarios, where Standard Model fields, such as the electromagnetic, live on a 
3+1 dimensional D3-brane while gravity propagates in the extra dimensions.

Here we consider a model where space-time has one timelike dimension and $d$ spatial 
dimensions and assume that Gauss law holds for the electric field,  
 i.e., the total electric flux on a closed surface 
in the $d$ dimensional space is proportional to the charge. This approach
is consistent with a scenario where the electromagnetic field is not constrained 
to a D3-brane.
We will study the Schr\"odinger equation for one electron atoms and 
search for  normalizable wave functions 
corresponding to bound states, i.e., negative energy eigenvalues. 
We will see that the $d=4$ case must be handled in a particular way.

\section{Schr\"odinger Equation}

The electric field produced by a static nucleus with
$Z$ protons of elementary charge $q$ in a space with $d$ non-compact spatial dimensions
depends on the distance $r$ as \cite{Zwiebach:2004tj}:
\begin{equation}
\vec E ( \vec r )\,=\, \frac{ \Gamma (d/2) }{ 2 \pi^{d/2} } \,
\frac{Z q}{r^{d-1}} \,\hat r\,\,.
\end{equation}

\bigskip
\noindent The corresponding static potential, taking as usual the potential 
at infinity to vanish, reads
\begin{equation}
\label{potential}
V ( r )\,=\, \frac{ \Gamma (d/2) }{ 2 \pi^{d/2} (d-2) }
\frac{ Z q}{r^{d-2}}\,\equiv \,
 \eta_d \,\frac{q}{ r^{d-2}}\,.
\end{equation}

\bigskip
Let us assume the existence of a stationary state (energy eigenstate)
 for an electron of mass $\mu$ with negative energy ${\cal E}$, 
and study the solutions of the corresponding Schr\"odinger equation 
\begin{equation}
\label{03}
\Big[  \frac{p^2}{2\mu} \,+\,U(r) \,\Big] 
\Phi_{\cal E} (\vec r )\,\,=\,\, {\cal E} \Phi_{\cal E} (\vec r )
\end{equation}
 
\noindent where $ U(r)\,=\,-\,q\,V(r)\,$ 

We are dealing with a central potential, so it is convenient to write
the eigenfunction as usual:
\begin{equation}
\Phi_{\cal E} (\vec r )\,=\,R_{\cal E} (r) \,Y_{\cal E} (\Omega )\,\,.
\end{equation}

The operator $p^2$ can be shown to act in this space with $d$ spatial dimensions
as \cite{Bars}

\begin{equation}
p^2 \,=\, p_r^2 \,+\, \frac{1}{r^2} \Big( L^2 \,+\, \frac{\hbar^2}{4} (d-1)(d-3) \Big)\,,
\end{equation}

\noindent where  the radial component $p_r$ acts on the radial part of the 
wave function as
\begin{equation}
\label{radial}
p_r^2 \,R (r)\,=\,- \hbar^2 r^{-\frac{d-1}{2}} \frac{d^2}{dr^2} 
\Big( r^{\frac{d-1}{2}} R(r) \Big)\,\,,
\end{equation}

\noindent while the angular momentum operator acts on the angular part  as\cite{Bars}
\begin{equation}
L^2\,Y_{\ell,m} (\Omega ) \,=\, \hbar^2 \ell ( \ell + 1 + d - 3)\,Y_{\ell,m} (\Omega )\,. 
\end{equation}

Equation (\ref{radial}) suggests   that we write the radial eigenfunctions as 
\begin{equation}
R_{{\cal E},\ell } (\vec r ) \,=\,r^{-\frac{d-1}{2}} 
f_{{\cal E},\ell} (r)   
\end{equation}

\noindent so that the radial equation takes the form:

\begin{equation}
\Big[ - \frac{d^2}{dr^2} \,+\, \frac{ \ell_d (\ell_d + 1)}{r^2}  
\,+\, \frac{2m}{\hbar^2} (\,U(r) - {\cal E}\,\,) \,\Big]\, f_{{\cal E},\ell} (r) 
\,=\,0 \,\,, 
\end{equation}
\vskip 0.4cm

\noindent where we introduced, for simplicity, 
\begin{equation}
\ell_d\,=\,\ell \,+\, (d - 3)/2   \,\,.
\end{equation}
This equation may be written in terms of dimensionless quantities 

\begin{equation}
\Big[  \frac{d^2}{d\rho^2} \,-\, \frac{ \ell_d (\ell_d + 1)}{\rho^2}  
\,+\, \frac{\eta_d}{\rho^{d-2}}\,-\,
\Big( \alpha^{\frac{3-d}{4-d}}\lambda_{\cal E} \,\Big)^2 \,\Big]\, 
f_{{\cal E},\ell} (\rho) \,=\,0  
\end{equation}

\vskip 0.4cm
\noindent where
$$ \alpha \,=\, \frac{mq^2}{\hbar^2}\,\,\,\,;\,\,\,\,
\rho \,=\, \alpha^{^{\frac{1}{4-d}}}\,r \,\,\,\,;\,\,\,
\lambda_{\cal E}^2 \,=\, - \frac{2\hbar^2}{mq^4} \,{\cal E}\,\,.
$$

At this point we note that this adimensionalization procedure,
well known for the three dimensional case, does not work for $d =4$. 

\section{ Case $d > 4$}

First, let us handle the case $d \ne 4$. 
Making the new substitution 

\begin{equation}
f_{{\cal E},\ell} (\rho) \,=\,exp\,\{ - \alpha^{\frac{3-d}{4-d}} 
\,\lambda_{\cal E} \rho \,\}\,\,
y_{{\cal E} , \ell} ( \rho )\,=\,0
\end{equation}

\noindent we obtain 
\begin{equation}
\label{14}
\Big[ \frac{ d^2}{d\rho^2} \,-\,2\,\alpha^{\frac{3-d}{4-d}} 
\,\lambda_{\cal E} \frac{d}{d \rho} \,+ \Big( \frac{\eta_d}{\rho^{d-2}} \,-\,
\frac{ \ell_d (\ell_d + 1)}{\rho^2}\Big) \Big] 
y_{{\cal E} , \ell} ( \rho )\,=\,0\,\,.
\end{equation}
   
Considering the fact that the wave function must be analytic, we can search 
for a power series solution: 
$$y_{{\cal E} , \ell} ( \rho )\,=\
\, \sum^\infty_{n = 0} c_n \,\rho^{n+s}\,.$$ 

\noindent  Equation (\ref{14}) gives us 
\begin{eqnarray}
c_0 &=& ... = c_{d-4} = 0
\nonumber\\
c_{n+d-3} &=& \frac{ \ell_d( \ell_d + 1) -
(n+s)(n+s-1)}{\eta_d} c_{n+1} \,+ \frac{ 2 \alpha^{\frac{3-d}{4-d}}
\lambda_{\cal E} (n+s)}{\eta_d}\,c_n
\end{eqnarray}

Hence, all the coefficients $c_n $ vanish and there is
no non-trivial solution
to $ y_{{\cal E} , \ell}$ in this case.  

\section{ Case $ d= 4$ }
The case d = 4 must be dealt with differently .
From (\ref{potential}) we have 
\begin{equation}
U ( r )\,=\,- \frac{ Z q^2 }{4 \pi^{2} r^2 }\,.
\end{equation}

\noindent The radial time-independent Scr\"odinger 
equation becomes 

\begin{equation}
\Big[ \frac{ d^2}{dr^2} \,+\,\frac{3}{r} \frac{d}{dr}\,-\,
\Big( \ell_4 (\ell_4 + 1) - \frac{3}{4} - \frac{m Z q^2}{2 \pi^2 \hbar^2}
\Big) \,\frac{1}{r} \,+\, \frac{2m{\cal E} }{\hbar^2} \Big] R (r)\,=\,0\,.
\end{equation}

\noindent Since we are only interested in bound states, it is convenient to introduce:
\begin{equation}
\epsilon^2 \,=\,- \frac{2 m {\cal E}}{\hbar^2}\,.
\end{equation}

\noindent Now, making the substitutions 
$$ \rho\,=\,\epsilon \,r\,\,,\,\,
 \,R ( \rho ) \equiv \frac{\epsilon}{\rho} g (\rho )\, \,,\,\,
 \zeta_{\ell} \,=\,1 + \ell_4 (\ell_4 + 1)\,-\,\frac{3}{4} - 
\frac{m Z q^2}{2 \pi^2 \hbar^2},$$

\noindent the radial equation takes the form

\begin{equation}
\Big[ \frac{ d^2}{d\rho^2} \,+\,\frac{1}{\rho} \frac{d}{d\rho}\,+\,
- \Big(   1 +  \frac{\zeta_{\ell}}{\rho^2} \Big) 
 \Big] g (\rho)\,=\,0\,.
\end{equation}
\vskip 0.4cm
\noindent The sign of the parameter $\zeta_{\ell} $ depends on $\ell_4\,=\,\ell\,+\,1/2$. 

Considering first the case of negative sign, we write $\zeta\,=\, - \nu^2 \,\,\,,$ 
with $\nu$  real and obtain 
\begin{equation}
\Big[ \frac{ d^2}{d\rho^2} \,+\,\frac{1}{\rho} \frac{d}{d\rho}\,+\,
- \Big(   1 -  \frac{\nu^2 }{\rho^2} \Big) 
 \Big] g (\rho)\,=\,0\,\,.
\end{equation}

Searching again for a series solution $ \sum^\infty_{n = 0} c_n \,\rho^{n+s}\,$ for
this equation we find the conditions
\begin{eqnarray}
(s^2 \,+\, \nu^2 ) c_0 &=&  0\nonumber\\
( (s+1)^2 + \nu^2 ) c_1 &=& 0 \nonumber\\
c_n &=& \frac{1}{ \nu^2 + (n+s)^2 }c_{n-2}
\end{eqnarray}

\noindent and, since $\nu^2$ is always non-zero, we conclude that there is no non-trivial
solution.

Now, considering $\zeta_{\ell} \ge 0\,$, the corresponding equation 
(using $\zeta\,=\, \nu^2 $) is

\begin{equation}
\Big[ \frac{ d^2}{d\rho^2} \,+\,\frac{1}{\rho} \frac{d}{d\rho}\,+\,
- \Big(   1 + \frac{\nu^2 }{\rho^2} \Big) 
 \Big] g (\rho)\,=\,0\,.
\end{equation}

\bigskip
\noindent This has the solutions $ I_\nu (\rho )\,$ and 
$K_\nu ( \rho )$ (modified Bessel functions).
But the first diverges at infinity, while the latter diverges at the origin. 
So, there is no physically acceptable (normalizable) bound state solution 
for the Schr\"odinger equation in this $d=4$ case.

\section{Conclusion}

We conclude that, assuming Gauss law still holds, there cannot be any
bound (negative energy) solution for the Schr\"odinger equation of the 
one-electron atom in higher non-compact dimensions. 
We point out that this result is consistent with that of its classical
gravitational analogue, the Kepler problem. 
In such systems, for a given $d \ge 4$, there is
only one stationary state, which disrupts under the smallest perturbation.
That is, the classical stationary states, 
corresponding to unstable ($d > 4$) or neutral ($d=4$)
equilibrium, must satisfy the conditions: 
$$ \Delta r \,=\,0\,\,\,\,;\,\,\,\,\Delta p_r\,=\,0\,\,.$$
\noindent In quantum mechanics the uncertainty relations forbid such a situation,
and the stationary states do not even occur.

Thus, the presence of atoms everywhere in our Universe  
places a restriction on the size of hypothetical extra dimensions.
They should have to be smaller than the atomic scale
unless the electromagnetic field does not propagate in these directions.
 
\acknowledgements The authors are partially supported by CNPq and FAPERJ.

\end{document}